\documentclass[prd,aps,twocolumn,a4paper,floatfix]{revtex4-2}

\usepackage{graphicx,psfrag,mathrsfs}
\usepackage{slashed}
\usepackage{mathrsfs}
\usepackage{amsmath,amsfonts,amssymb,amsthm}
\usepackage{hyperref}
\hypersetup{breaklinks=true}
\usepackage{url}
\usepackage{comment,cancel}
\usepackage{accents}
\usepackage{ulem}
\usepackage{xcolor,bbold}

\makeatletter
\newcommand{\sbullet}{%
  \hbox{\fontfamily{lmr}\fontsize{.4\dimexpr(\f@size pt)}{0}
    \selectfont\textbullet}}

\makeatother

\def\mbn{\overset{\bullet}{\nabla}}

\def\sD{\slashed{D}}

\def\sD{\slashed{D}} 
\def\sg{\slashed{g}}
\def\psib{\ul{\psi}}

\def\Rc{\mathring{R}}

\def\p{\partial}

\def\non{\nonumber}
\def\ul{\underline}

\newtheorem{thm}{Theorem}

\allowdisplaybreaks

\begin{document}

\title{Energy estimates for the good-bad-ugly model}

\author{Miguel Duarte}

\affiliation{
  CENTRA, Departamento de F\'isica, Instituto Superior
  T\'ecnico IST, Universidade de Lisboa UL, Avenida Rovisco Pais 1,
  1049 Lisboa, Portugal,
}

\begin{abstract}
We establish a relationship between the equations that constitute the so-called \textit{good-bad-ugly model}, whose nonlinearities are known to mimic those present in the Einstein field equations in generalized harmonic gauge. This relationship between ugly fields and good and bad ones stems from the fact that one can write the equation for the rescaled derivative of an ugly along an incoming null direction as a good or a bad equation depending on whether there are source terms or not. This provides a new interpretation of the logarithms of the radial coordinate that show up in expansions of solutions to ugly equations near null infinity. This furthermore allows us to use the Klainerman-Sobolev inequality for the standard wave equation on Cauchy slices to show uniform boundedness for the ugly equation. In the second part of this paper we perform a first order reduction of the ugly equation with given sources in flat space and we radially compactify the coordinates in order to show an energy estimate for that equation on hyperboloidal slices. This result is an important first step towards establishing energy estimates for the hyperboloidal initial value problem of the first order compactified Einstein field equations in generalized harmonic gauge.
\end{abstract}

\maketitle

\section{Introduction}

As is well known, gravitational radiation is non-localizable and hence only well defined at null infinity. For this reason, there have been various attempts to include null infinity in the computational domain in order to avoid having to evaluate waveforms at large radius and extrapolate the rest or the way. The latter approach has been shown to work but, from a mathematical point of view, the former would be undeniably better. Using Penrose's idea to bring null infinity to a finite coordinate radius~\cite{Pen63} and Friedrich's conformal Einstein field equations (CEFE), H\"ubner~\cite{Hub99,Hub01} and Frauendiener~\cite{DouFra16} have used conformal compactification to solve this problem. The use of CEFEs has brought about important results in recent years, but it is not yet clear how to lift over standard methods to treat the strong field region of spacetimes of interest.

Other approaches include using Cauchy-characteristic matching~\cite{Win05} and assigning initial data on a hyperboloidal slice instead of on a more standard Cauchy slice~\cite{BucPfeBar09,BarSarBuc11,MonRin08,RinMon13}. Hyperboloidal slices are hypersurfaces that are everywhere spacelike but nevertheless intersect null infinity. These are not Cauchy hypersurfaces as their domain of dependence does not cover the entire spacetime. Although having already shown remarkable results, using standard formulations of General Relativity (GR) on hyperboloidal slices has the particularity that it seems to inevitably give rise to formally singular terms, which are regular but that relies on prior knowledge about the behaviour of evolved fields near null infinity and hence create problems in numerical evolutions. These terms have been successfully treated through the use of L'H\^opital's rule at null infinity. See~\cite{Zen07,VanHusHil14,VanHus14,Van15} for a treatment of the spherically symmetric case. 

An important proposal that aims to provide an alternative to the use of the CEFE while insisting on the inclusion of null infinity in the computational domain in the hyperboloidal setting is the dual-frame approach~\cite{Hil15}. This is essentially a way to decouple the coordinates from the tensor basis and carefully choose each of them so that the Einstein field equations (EFE) can be written in generalized harmonic gauge (GHG) and then solved with hyperboloidal coordinates~\cite{HilHarBug16,GasHil18,GasGauHil19,GauVanHil21}. The former choice allows the EFE to be written as a set of inhomogeneous wave equations, while the latter allows us to bring null infinity to a finite coordinate radius through hyperboloidal compactification. Essential aspects of this approach are the \textit{coordinate light speed condition}, the requirement that radial coordinate light speeds have a certain decay at null infinity, and the \textit{weak null condition}~\cite{GasHil18,LinRod03}, a property of the non-linearities of quasi-linear wave equations that is expected to be sufficient to establish small data global existence. An important step towards proving this result in full generality has been the work of Keir~\cite{Kei17} which shows that this is implied by the \textit{hierarchical weak null condition}. Additionally, in an attempt to drop the assumptions on the structure of the nonlinearities, small data global existence has been shown for cases where solutions are bounded and stable against rapidly decaying perturbations~\cite{Kei19}.

It has been shown that the simplest choices of initial data in the dual-frame approach give rise to logarithmically (formally) divergent terms, which are problematic from the point of view of numerical evolutions. To solve this problem, it was found that, for a toy model dubbed \textit{good-bad-ugly} whose non-linearities mimic those present in the EFE in GHG, these logarithms can be explicitly regularized at first order through a non-linear change of variables~\cite{GasGauHil19}. In parallel, a heuristic method to find solutions to the \textit{good-bad-ugly} system near null infinity was laid out, building on H\"ormander's idea of asymptotic systems~\cite{DuaFenGasHil21}. Similar results were found for a generalized version of this toy model that included wave operators built from asymptotically flat metrics, as well as inhomogeneities that depend on the evolved fields but do not contribute to leading order, the so-called \textit{stratified null forms}. These findings led to the development of a heuristic method that is able to predict the logarithms that may appear in solutions to the EFE in GHG and to make those logarithms disappear with constraint addition and a careful choice of gauge. This was used to investigate whether or not the EFE satisfy the \textit{peeling} property, a requirement on the decay of the components of the Weyl tensor that ensures smooth null infinity~\cite{DuaFenGas22}. Finally, in~\cite{DuaFenGasHil22a}, the authors make use of this method to find a first order reduction of the EFE in GHG that is radially compactified and ready to be implemmented numerically.

Adapting H\"ormander's method, recent work has shown heuristically that, for a large class of regular initial data, even solutions to the wave equation allow for the appearance logarithms at every order~\cite{DuaFenGasHil25}. This then results in an obstruction to the peeling property.

Given the fact that the results discussed above are heuristic, an important next step is to find estimates for the system of equations in question. This work can be seen as a modest first step towards that goal. Taking an ugly equation in Minkowski spacetime with given source terms, we show an important relationship between ugly fields and good and bad ones that stems from the fact that one can write the equation for the rescaled derivative of an ugly along an incoming null direction as a good or a bad equation depending on the given sources. This is done by adapting the asymptotic systems method and it allows us to find energy estimates for an ugly equation on Cauchy slices by using the Klainerman-Sobolev inequality, a result that applies to the much more extensively studied good equation. The second part of this paper is dedicated to finding energy estimates for that equation on hyperboloidal slices. We expect these results to be an important stepping stone in finding energy estimates for the hyperboloidal initial value problem of the first order compatified EFE in GHG.

\section{Setup}
In the following we describe the geometric setup on which the present paper is built, we introduce the type of equations we will be working with and mention some relevant results obtained in previous work.
\subsection{Representation of the metric} We define a flat metric $\eta_{ab}$ with Levi-Civita connection $\nabla$. As in this work we will deal exclusively with flat spacetimes, indices will be raised and lowered with $\eta_{ab}$ except when mentioned otherwise. We introduce a global inertial Cartesian coordinate system~$X^{\ul{\alpha}}=(T,X^{\ul{i}})$. Let~$\p_{\ul{\alpha}}$ and~$dX^{\ul{\alpha}}$ be the
corresponding vector and co-vector bases. The flat covariant derivative
associated to $X^{\ul{\alpha}}$ on a flat background coincides with $\nabla$. Additionally we define the shell
coordinates~$X^{\ul{\alpha}'}=(T',X^{\ul{i}'})=(T,R,\theta^A)$, where
the radial coordinate $R$ is related to $X^{\ul{i}}$ as~$R^2=(X^{\ul{1}})^2+(X^{\ul{2}})^2+(X^{\ul{3}})^2$. Let~$\p_{\ul{\alpha}'}$
and~$dX^{\ul{\alpha}'}$ be the corresponding vector and co-vector
bases. Shell coordinates have an associated flat covariant
derivative~$\mbn$, with the defining property that $\mbn_b\p_{\ul{\alpha}'}^a=0$. We define outgoing and incoming null vector fields in the shell coordinate basis as,
\begin{align}\label{eq:psiinshellchart-Est}
  \psi^a&=\p_T^a+\p_R^a\,,\nonumber\\
  \ul{\psi}^a&=\p_T^a-\p_R^a\,,
\end{align}
and we write the metric with upstairs indices as,
\begin{align}\label{eq:metricrepresentation-Est}
\eta^{ab}=-\psi^{(a}\ul{\psi}^{b)}+\slashed{\eta}^{ab} \,,
\end{align}
where $\slashed{\eta}^{AB}$ is exactly $\eta^{AB}$. We define the derivative~$\sD_A$ for scalar functions $\phi$,
\begin{align}\label{sDDef}
\sD_A \phi:= \slashed{\eta}_{A}{}^{b}\mbn_b \phi\,.
\end{align}
For simplicity of the expressions, the index associated with this derivative and only that index will be raised with $\slashed{\eta}^{ab}$. We also define the vector field~$T^a:=\p_T^a$ and denote the covariant
derivative in the direction of~$T^a$ as~$\nabla_T$.
\subsection{Our model and assumptions} The equations we are concerned with here are,
\begin{align}\label{gbu-Est}
	&\square g = F_g\,,\non\\
	&\square b = (\nabla_T g)^2+F_b\,,\non\\
	&\square u = \frac{2p}{R}\nabla_T u+F_u\,,
\end{align}
where $\square$ is the standard wave operator built from the flat metric $\eta_{ab}$, $p$ is a natural number that we will leave free for now and $F_\phi$ are given functions of order $O(R^{-3})$. The letters chosen to denote the evolved fields stand for \textit{good}, \textit{bad} and \textit{ugly}, respectively, which justifies the name of the model. Throughout this work we will use these names to refer to fields that satisfy equations of the type~\eqref{gbu-Est}. In line with H\"ormander's asymptotic systems method, we assume that derivatives of evolved fields along $\psib^a$ or $\p_T^a$ (bad derivatives) keep their decay near null infinity, whereas derivatives along $\psi^a$ (good derivatives) improve it.
\subsection{The good and the ugly} It was seen in~\cite{DuaFenGas22} that if we generalize equations~\eqref{gbu-Est} to have wave operators built from metrics that are asymptotically flat and dependent on the evolved fields themselves, and include inhomogeneities on the right-hand side that are also dependent on the fields but do not contribute to leading order in $R$, then that type of equations encompasses the EFE in GHG upon a specific choice of variables. In the same work, the authors show that a careful choice of gauge and constraint addition allows us to write the EFE as a system of only 2 goods, 8 uglies and a gauge driver (which behaves asymptotically as a good field). This, together with the fact that the standard wave equation has been studied extensively, motivates a special focus on the ugly equation.
\section{Energy estimates on Cauchy slices}\label{UgliesAreGoods-Est}
Let us consider the single equation,
\begin{align}\label{uglyEst}
	\square \phi = \frac{2p}{R}\nabla_T \phi + F\,,
\end{align}
where $F$ is a given function of order $O(R^{-3})$ and hence could be expected not to influence the leading order asymptotics of the equation near null infinity. We name the evolved field $\phi$ instead of $u$ for ugly because this model can easily be turned into a good equation just by setting $p=0$. We will now be working with only one equation for simplicity, but results like the ones we aim to show are easy to generalize for systems of equations. Note that there are two major simplifications in this toy model as compared to the ugly fields in the EFE in GHG, namely the fact that $\square$ is now built from a flat metric and the choice of working with a given $F$ instead of a combination of stratified null forms that could depend upon the evolved fields~\cite{DuaFenGasHil21}. In the first part of this section we want to show that there is a very close relationship between the ugly, the bad and the good equations, which can be used to find energy estimates for an ugly field from estimates for the much better studied good field. This we will do in the second part of this section for the special case $p=1$ and $F=0$.
\subsection{The standard ugly case} 
We assume $\phi$ satisfies the ugly equation~\eqref{uglyEst} with $p=1$,
\begin{align}\label{uglyEst2}
	\square \phi = \frac{2}{R}\nabla_T \phi + F\,.
\end{align}
This simplification is done because the relation we wish to show here is simpler for this particular case and we leave an analysis of the case where $p\neq 1$ for the next subsection. Additionally, we assume that the solutions to this equation satisfies,
\begin{align}
	u=o^+(1)\,,
\end{align}
near null infinity, where the notation~$f=o^+(h)$ means,
\begin{align}\label{deflittleo-Est}
  \exists \epsilon>0 :
  \lim_{R\rightarrow\infty} \frac{f}{hR^{-\epsilon}}=0\,,
\end{align}
which can be informally stated as \textit{$f$ falls-off faster than~$h^{1+\epsilon}$ as $R$ goes to infinity}, which is a faster decay than~$f=o(h)$. Note that we will only use this assumption in this subsection and the next. This is because it is needed to establish results using the asymptotic systems method, but would make no sense to assume when showing energy estimates, since that would essentially mean assuming what we are trying to prove. If~\eqref{uglyEst2} holds then,
\begin{align}\label{EqRpsibPhi}
	\square(R\nabla_{\psib}\phi) = \frac{1}{R}\nabla_{\psib}(R^2F)\,,
\end{align}
also holds. Considering that $F$ is of order $O(R^{-3})$ and assuming that, in the worst case scenario, its derivative along the incoming null vector field $\psib^a$ does not improve its decay, then equation~\eqref{EqRpsibPhi} is a bad equation for $R\nabla_{\psib}\phi$ that becomes good if $F=0$ or if a bad derivative of $F$ improves its decay. For simplicity let us momentarily consider that to be the case. Note that up to this point we have not yet made any assumptions on initial data. However, it is interesting to see how this relationship between different types of fields plays out in solutions found in the literature. In~\cite{DuaFenGasHil25} the authors found asymptotic solutions of the good-bad-ugly model near null infinity using initial data of the following class: Let $\Sigma_0 $ be a Cauchy slice defined by the condition $T=0$. Initial data on $\Sigma_0 $ is given by,
\begin{align}\label{ID}
	\phi\rvert_{\Sigma_0 } =
        \sum_{n=1}^{\infty}\frac{\bar{M}_{\phi,n}}{R^n}\,,\quad
        \nabla_T\phi\rvert_{\Sigma_0 }=\sum_{n=1}^{\infty}\frac{M_{\phi,n}}{R^{n+1}}\,,
\end{align}
where both $M_{\phi,n}$ and $\bar{M}_{\phi,n}$ are functions of only the angles in the Shell coordinate system, and $\phi$ can mean any of evolved fields in~\eqref{gbu-Est}. For the sake of comparison with earlier work we will consider initial data of the type~\eqref{ID} for this subsection and the next. However, more restrictive assumptions will be placed later on when we use the Klainerman-Sobolev inequality to establish a uniform boundedness result for ugly fields. Taking the functional form of good equations from~\cite{DuaFenGasHil25}, what~\eqref{EqRpsibPhi} tells us is that the variable we get by hitting $\phi$ with a $\psib^a$ derivative and rescaling it by $R$ behaves as,
\begin{align}\label{goodPolyhomo}
	R\nabla_{\psib}\phi = \sum_{n=1}^\infty \frac{\Theta_{n,0}(\psi^*)+\Theta_{n,1}(\psi^*)\log R}{R^n}\,,
\end{align}
where $\Theta_{n,0}(\psi^*)$ and $\Theta_{n,1}(\psi^*)$ are scalar functions that do not vary along integral curves of the outgoing null vector field $\psi^a$. Again, from~\cite{DuaFenGasHil25}, we know that ugly equations with $p=1$ admit asymptotic solutions of the form,
\begin{align}\label{uglyPolyhomo-Est}
	u =\frac{\mathcal{U}_{1,0}(\theta^A)}{R}+\sum_{n=2}^{\infty}\sum_{k=0}^{N_n^u}
  \frac{(\log R)^k \mathcal{U}_{n,k}(\psi^*)}{R^n}\,,
\end{align}
for some natural number $N_n^u$. Due to this deep relationship between goods and uglies, hitting~\eqref{uglyPolyhomo-Est} with a derivative along $\psib^a$ and rescaling it by $R$ should yield an expression of the type~\eqref{goodPolyhomo}. However, that is the case only if $N_n^u=1$. This result seems to suggest that either $p\neq 1$ introduces higher powers of logs or the hypothesis proved by induction in~\cite{DuaFenGasHil25} was not the sharpest it could have been. Namely, uglies in flat spacetimes with $p=1$ seem to only allow for logs appearing linearly at any given order beyond the first (answering this question is beyond the scope of this work). Let us now consider that $F\neq 0$, in which case the variable $R\nabla_{\psib}\phi$ behaves a bad field. From~\cite{DuaFenGasHil25}, bad fields are expected to behave asymptotically as,
\begin{align}\label{badPolyhomo}
  &b =  \sum_{n=1}^{\infty}\sum_{k=0}^{n} \frac{(\log R)^k \mathcal{B}_{n,k}(\psi^*)}{R^n}\,.
\end{align}
Once more hitting~\eqref{uglyPolyhomo-Est} with a derivative along the incoming null direction $\psib^a$ and rescaling it produces an expression consistent with~\eqref{badPolyhomo}.
\subsection{Pushing logs down in the expansion} 
In~\cite{DuaFenGasHil25} the authors showed that introducing a natural number $p$ in the ugly model as in~\eqref{uglyEst} suppresses the second order logarithm up to order $p+1$ in $R^{-1}$. It is interesting to see how this property manifests in the equation for the rescaled $\psib^a$ derivative of the field. In the last subsection we saw that there is a close relationship between ugly fields with $p=1$ and good or bad fields depending on whether $F$ is set to zero or not. Predictably, that relationship does not hold for higher $p$, but there is a close connection between uglies of consecutive $p$. In order to see that, let us consider the model~\eqref{uglyEst} again and, instead of finding an equation for the variable $R\nabla_{\psib}\phi$, we find one for $\nabla_T\Phi :=R\nabla_T \phi$,
\begin{align}\label{CommutatorWithp-Est}
	\square(\nabla_T\Phi) = \frac{2(p-1)}{R}\nabla_T(\nabla_T\Phi) + R\nabla_T F + \frac{2}{R}\nabla_\psi(\nabla_T\Phi)\,.
\end{align}
We choose to work with this variable because the commutator of $\nabla_T$ with $\square$ in this particular model is cleaner. We now place the usual assumptions that good derivatives improve the decay of solutions near null infinity, whereas bad ones do not, and apply them to $(\nabla_T\Phi)$. The left-hand side and the first two terms on the right-hand side contribute to the leading order of the solution, but the third one does not. Equation~\eqref{CommutatorWithp-Est} then does not fit our definitions of good, bad or ugly fields so long as $F\neq0$. However, if we set $F=0$,~\eqref{CommutatorWithp-Est} becomes an ugly equation with $p$ shifted by $1$ and a so-called stratified null form, $\frac{2}{R}\nabla_\psi(\nabla_T\Phi)$, which does not affect the leading order behavior near null infinity (see~\cite{DuaFenGasHil21} for a detailed discussion of these terms). We can integrate~\eqref{CommutatorWithp-Est} in exactly the same way as was done in~\cite{DuaFenGasHil25} to get,
\begin{align}\label{asymp-Est}
	&(R\nabla_T\Phi)\simeq \frac{1}{2R^{p-1}}\int_{-R}^u\int_{-u'}^R
        R'^{p}\nabla_T FdR'du'\\ &+
        \frac{1}{R^{p-1}}\int_{-R}^u\left[\nabla_T (R^{p}\nabla_T\Phi)\right]\rvert_{R\rightarrow-u'}du'+(R\nabla_T\Phi)\rvert_{u\rightarrow -R}\,.\non
\end{align}
If we consider initial data of the type~\eqref{ID} for our new field $(\nabla_T\Phi)$, then we have,
\begin{align}
	&(R\nabla_T\Phi)_{\Sigma_0 }\simeq \bar{M}_{(\nabla_T\Phi),1}\,,\non\\
	&\nabla_T(R\nabla_T\Phi)_{\Sigma_0 }\simeq \frac{M_{(\nabla_T\Phi),1}}{R}.
\end{align}
Plugging this into~\eqref{asymp-Est} we get,
\begin{align}\label{asymp-Est2}
	(R\nabla_T\Phi)&\simeq \frac{1}{2R^{p-1}}\int_{-R}^u\int_{-u'}^R
        R'^{p}\nabla_T FdR'du'\\ 
        &+\frac{M_{(\nabla_T\Phi),1}}{p-1}\left[1-\frac{(-u)^{p-1}}{R^p}\right]+\bar{M}_{(\nabla_T\Phi),1}\,,\non
\end{align}
for $p>1$. The only term in~\eqref{asymp-Est2} that can generate a logarithm is the first one on the right-hand side. However, it can only do that at order $R^{-1}$, so we can see through this method that increasing $p$ effectively pushes logs down in the expansion.
\subsection{Finding estimates in $L^2$} 
\begin{figure}[t!]
\centering
\includegraphics[width=0.45\textwidth]{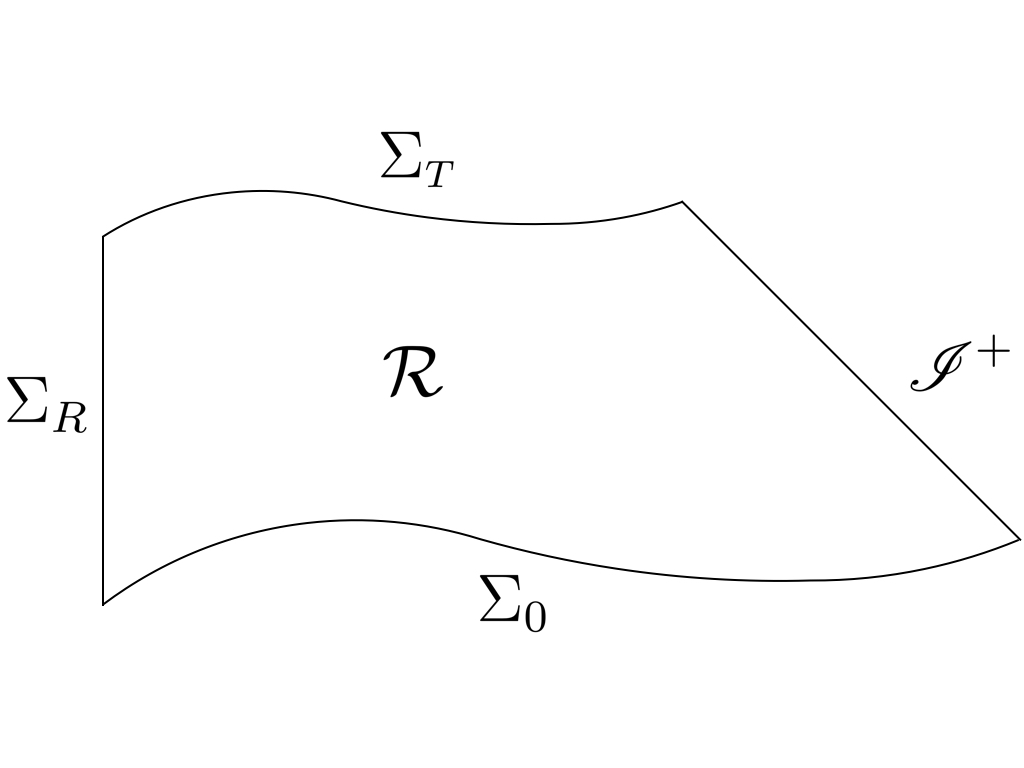}
\caption{A schematic of our geometric setup.}
  \label{fig:Cauchy}
\end{figure}

The most interesting use for the fact that equations for uglies can be written as bad equations or good ones for the rescaled $\psib^a$ derivatives of the original fields is that, by turning ugly equations into goods, we can then use the energy estimates that have already been proven for the much more extensively studied standard wave equation on Cauchy slices without the need of coming up with entirely new theorems which apply only to uglies. Naturally, from the discussion above, this is most direct in flat space and in the case where $p=1$ and $F=0$. In the following we want to establish energy estimates in $L^2$ for the standard wave equation and its derivatives that will allow us to use the Klainerman-Sobolev inequality to obtain $L^\infty$ estimates. Then, using the results above, namely the relationship between uglies and goods, we can find an estimate in $L^\infty$ for solutions to the ugly equation. Let $\mathcal{R}$ be a region of spacetime comprised between the Cauchy hypersurfaces $\Sigma_0$ and $\Sigma_T$, defined by taking constant values of the time coordinate $T$, the boundary of a world-tube $\Sigma_R$ and null infinity $\mathscr{I}^+$ as shown on Figure~\ref{fig:Cauchy}.
Let the field $\phi_1$ satisfy,
\begin{align}\label{wave}
	\square \phi_1 = 0\,.
\end{align}
We define the energy norm of $\phi_1$ on a Cauchy slice $\Sigma_T$ as,
\begin{align}
	E^2(T) = \frac{1}{2}\int_{\Sigma_T}\left[ (\p_{T'}\phi_1)^2 + (\p_{\ul{i}}\phi_1)(\p^{\ul{i}}\phi_1)\right]dX\,,
\end{align}
where $\p^{\ul{i}}=\delta^{\ul{ij}}\p_{\ul{j}}$ and $dX$ is defined as,
\begin{align}
	dX = dX^1dX^2dX^3\,.
\end{align}
Multiplying equation~\eqref{wave} by $\p_T\phi_1$ and integrating in $\mathcal{R}$ gives,
\begin{align}
	\int_\mathcal{R} \p_{T'}(\p_{T'}^2\phi_1-\p_{\ul{i}}\p^{\ul{i}}\phi_1)dV = 0\,,
\end{align}
where,
\begin{align}
	dV = dT'dX\,.
\end{align}
Integrating by parts and discarding the boundary terms we get,
\begin{align}
	E^2(T)=E^2(0)\,.
\end{align}
Therefore, the energy is conserved and the $L_2$ norms of the first derivatives of $\phi_1$ are bounded. 

\subsection{Finding estimates in $L^\infty$}
Throughout the rest of this section we will consider initial data of compact support. To enunciate the Klainerman-Sobolev inequality, we need to define some derivatives:
\begin{itemize}
	\item Translations: Cartesian coordinate vector fields $\p_{\ul{\alpha}}$;
	\item Spatial rotations: $\Omega_{\ul{ij}}:=\eta_{\ul{ik}}X^{\ul{k}}\p_{\ul{j}}-\eta_{\ul{jk}}X^{\ul{k}}\p_{\ul{i}}$;
	\item Lorenz boosts: $\Omega_{T\ul{i}}:=\eta_{\ul{ik}}X^{\ul{k}}\p_{T}-T\p_{\ul{i}}$;
	\item Dilations $S:=T\p_T+X^{\ul{i}}\p_{\ul{i}}$.
\end{itemize}
We order these derivatives in some arbitrary way and name them $\Gamma_q$, where $q$ is a natural number that satisfies $1\leq q\leq N$, where $N$ is the total number of distinct derivatives defined. Furthermore we define the multi-index notation as follows. Given a multi-index $I=(I_1,...,I_N)$, where $1\leq I_q\leq N$, we define,
\begin{align}
	\Gamma^I := \Gamma_{I_1}\Gamma_{I_1}...\Gamma_{I_N}\,.
\end{align}
The Klainerman-Sobolev inequality~\cite{Sog95} goes as follows,
\begin{thm} 
Let $\phi_1\in C^{\infty}([0,\infty)\times \mathbb{R}^3)$ such that $\phi_1(T)\in\mathbb{R}^3$ is a function of compact support for any $T\geq 0$. Then the following estimate holds for each $T\geq 0$ and $X^{\ul{i}}\in \mathbb{R}^3$,
\begin{align}\label{contentKlainerman}
	(1+T+R)(1+|T-R|&)^{1/2}|\phi_1(T,X)| \non\\
	&\leq \sum_{|I|\leq \frac{5}{2}} ||\Gamma^I \phi_1(T)||_{L^2}\,.
\end{align}
\end{thm}
We can apply the theorem directly to our field $\phi_1$, but that alone does not guarantee that the left-hand side of~\eqref{contentKlainerman} is bounded. We still need to provide bounds for the right-hand side. A straightforward calculation shows that all the derivatives $\Gamma_q$ defined above commute with the flat space wave operator $\square$, which means that,
\begin{align}
	\square (\Gamma_q \phi_1) = 0\,
\end{align}
As $\Gamma_q \phi_1$ satisfies the wave equation, exactly the same energy method can be applied to it to show that its derivatives are bounded as well. This provides the bounds we needed on the right-hand side of~\eqref{contentKlainerman} and so the inequality guarantees that the left-hand side,
\begin{align}\label{LHS_KS}
	(1+T+R)(1+|T-R|&)^{1/2}|\phi_1(T,X)|\,,
\end{align}
is bounded. It is interesting see how this relates to what we know about ugly fields close to null infinity. As mentioned above, earlier work~\cite{GasGauHil19,DuaFenGasHil21} around this field showed that it behaved slightly better than solutions to the standard wave equation near null infinity. Namely, that $\psib^a$ derivatives of it gained one order in $R^{-1}$ as compared to the decay of the field itself, which is not true for good fields. Let us take a closer look at the bounded quantity~\eqref{LHS_KS}. As we go out to null infinity along integral curves of the outgoing null vector field $\psi^a$, the first factor, $(1+T+R)$, grows whereas the second, $(1+T-R)$ remains constant. Now, if we assumed that $\phi_1(T,X)$ is a solution to the good equation, then $\phi$ is a solution to an ugly one with $p=1$ and $F=0$ as long as $\phi_1=R\nabla_{\psib}\phi$. Since~\eqref{LHS_KS} is bounded, then $\nabla_{\psib}\phi$ must decay at least like,
\begin{align}
	|\nabla_{\psib}\phi(T,X)|\sim\frac{1}{R(1+T+R)}\,.
\end{align}
In other words, the $\psib^a$ derivative of an ugly field decays one order faster than that of a good field. This result, which had only been reached through heuristic methods, is now proven rigorously through energy estimates and the use of the Klainerman-Sobolev inequality, for initial data of compact support.
\section{Radially compactified ugly}\label{Compactification-Est}
In~\cite{DuaFenGasHil22a} the authors wrote the EFE in GHG as a system of goods, bads and uglies, and carefully chose a gauge and constraint addition such that the 10 independent metric components turned into 2 goods and 8 uglies. Then, they performed a first order reduction of the equations and a coordinate change in order to bring null infinity to a finite coordinate distance. The result was a system of 55 first order differential equations for 55 unkowns. The goal here is to use energy methods to find the decay of solutions to the good-ugly model while avoiding heuristic arguments. This is done with two major simplifications, namely, the metric is taken to be the Minkowski metric, the sources on the right-hand side are considered to be given functions of the coordinates and, for simplicity, we only work with one equation for the field $\phi$. Naturally, these simplifications rule out GR, but we believe that working out the details of energy estimates for these equations is a crucial step towards finding estimates for the EFE in full generality.
\subsection{First order reduction}
In order to make a first order reduction we treat each derivative of the original field $\phi$ as a variable in its own right and then choose a sufficient number of equations so that we have one for each reduction variable. Let $\phi_\psi$, $\phi_{\psib}$ and $\phi_A$ denote the derivatives $\nabla_\psi\phi$, $\nabla_{\psib}\phi$ and $\sD_A\phi$, respectively. This means we have now 5 independent variables for which we must pick 5 independent first order equations. As we will see, there is a certain freedom in the equations we use. One of them will necessarily be~\eqref{uglyEst}, since it represents goods and uglies depending on the integer $p$, which takes the form,
\begin{align}\label{uEqLong}
  \frac{1}{R^{p+1}}\nabla_\psi\left[R^{p+1}\phi_{\psib}\right]
  - \sD^A\phi_A = -\frac{p-1}{R}\phi_{\psi}+\frac{\cot \theta}{R^2}\phi_{\theta}-F.
\end{align}
The others are not as obvious. We take the torsion-free condition $\mbn_a\nabla_b\phi=\mbn_b\nabla_a\phi$  and contract it with the vectors $\psi^a$ and $\psib^a$, and $T^a$ and $\sg_A{}^a$. Here we use the shell covariant derivative in line with~\cite{DuaFenGasHil22a}. The first combination of contractions yields,
\begin{align}
  \psi^a\psib^b\mbn_a\nabla_b\phi&=\psi^a\psib^b\mbn_b\nabla_a\phi\,,
\end{align}
which means,
\begin{align}
  \nabla_\psi\phi_{\psib}&=\nabla_{\psib}\phi_{\psi}\,,
\end{align}
but the left-hand side is already given by~\eqref{uEqLong}, so we can write,
\begin{align}\label{notorsionNorescale1}
  \nabla_{\psib}\phi_{\psi} = &\sD^A\phi_A -\frac{p+1}{R}\phi_{\psib} -\frac{p-1}{R}\phi_\psi+\frac{\cot \theta}{R^2}\phi_\theta-F\,.
\end{align}
Contracting the torsion-free condition with the second combination of vectors, $T^a$ and $\sg_A{}^a$ we get,
\begin{align}\label{notorsionNorescale2}
  &2\nabla_{\psib}\phi_{A}-2\nabla_{\psi}\phi_{A}-\sD_A\phi_{\psib}
  -\sD_A\phi_{\psi} =0\,.
\end{align}
We now have one equation for each of the reduction variable, leaving us only with the task of choosing an equation that will relate these to the original variable $\phi$. We pick the equation,
\begin{align}\label{redConstraintNorescale}
  \phi_{,\psib} = \nabla_{\psib}\phi\,,
\end{align}
to serve that purpose and treat,
\begin{align}\label{RedConstraints}
  \phi_{,\psi} = \nabla_\psi\phi\,,\quad\phi_{,A}
  = \sD_A\phi\,,
\end{align}
as constraints that one would need to make sure are satisfied everywhere when evolving the system numerically. Note that equation~\eqref{redConstraintNorescale} is decoupled from the rest and so our results can be shown without the complication of adding a fifth equation to the system. For this reason we will work only with 4 equations,~\eqref{uEqLong},~\eqref{notorsionNorescale1} and~\eqref{notorsionNorescale2}. The next step is to rescale the chosen variables by as many powers of $R$ as possible and finding equations for the resulting variables. Once more we do this in the same way as was done in~\cite{DuaFenGasHil22a},
\begin{align}\label{rescaling-Est}
	&\Phi:=R\phi\quad,\quad\Phi_\psi:=R\nabla_\psi\Phi\,,\non\\
	&\Phi_{\psib}:=\nabla_{\psib}\Phi\quad,\quad\Phi_A:=\sD_A\Phi\,.
\end{align}
This choice of equations and rescalings gives the following system,
\begin{align}\label{EqsRescaled-Est}
	&\frac{1}{R^{p+1}}\nabla_\psi\left[R^p\Phi_{\psib}\right]-\frac{1}{R}\sD^A\Phi_A = \frac{\cot\theta}{R^3}\Phi_\theta-\frac{p}{R^3}\Phi_\psi-F\,,\non\\
	&\frac{1}{R^2}\nabla_{\psib}\Phi_{\psi} -\frac{1}{R}\sD^A\Phi_A+ \frac{p}{R^2}\Phi_{\psib} = \frac{\cot\theta}{R^3}\Phi_\theta-\frac{p+1}{R^3}\Phi_\psi-F\,,\non\\
	&2\nabla_T\Phi_A -\sD_A\Phi_{\psib} - \frac{1}{R}\sD_A\Phi_{\psi} = 0\,.
\end{align}
According to~\cite{DuaFenGasHil22a}, the next step would be to change the coordinates to a set that makes use of the areal radius $\Rc$ instead of the standard radial coordinate $R$, but in flat space these two radii coincide, so we are now ready to radially compactify the coordinates.	
\subsection{Radial compactification}
Compactification is a method used to bring null infinity to a finite value of the radial coordinate. On a flat background, this is done by considering the coordinate system $(t,r,\bar{\theta}^A)$ which is related to the previous one by,
\begin{align}\label{3rdCoords-Est}
	T=t+H(R(r)),\quad R= R(r),\quad\bar{\theta}^A=\theta^A\,,
\end{align}
where $H(R(r))$ and $R(r)$ are called height and compression functions, respectively. We define a set of vector fields associated to~\eqref{3rdCoords-Est},
\begin{align}
  \xi^a&=\p_{ t}^a+\mathcal{C}^{ r}_+\p_{ r}^a\,,\nonumber\\
  \ul{\xi}^a&=\p_{ t}^a+\mathcal{C}^{ r}_-\p_{ r}^a\,,
\end{align}
where the coordinate light speeds $\mathcal{C}^{ r}_\pm$ are fixed by requiring that the vectors $\xi^a$ and $\ul{\xi}^a$ are null with respect to the metric $\eta_{ab}$. Naturally, these quantities depend upon the choice of height and compression functions, namely,
\begin{align}\label{H'-Est}
	H'=1-\frac{1}{R'\mathcal{C}^{ r}_+}\,,
\end{align}
where $H':=\p_{R}H$ and $R':=\p_{r}R$. Let us then make that choice by imposing that $\mathcal{C}^{ r}_+=1$. We get,
\begin{align}\label{H'2-Est}
	H'=1-\frac{1}{R'}\,.
\end{align}
We do not make a strict choice for the compression function, we simply require that it satisfies,
\begin{align}\label{RprimeAssump-Est}
	R'(r)\simeq R^n\,,
\end{align}
where $1<n\leq 2$. The lower bound on $n$ is necessary to make $r$ approach a finite value as $R$ goes to null infinity, whereas $0<n\leq 2$ is required for numerical stability, as discussed in~\cite{CalGunHil05}. To find the compactified form of the equations we could now do it directly by relating the null vectors $\psi^a$ and $\psib^a$ with $\xi^a$ and $\ul{\xi}^a$. However it pays off to do it with matrices. We therefore define a vector $\mathbf{v}^a$ whose entries are each of the reduction variables,
\begin{align}
	\mathbf{v} = (\Phi_{\psib},\Phi_\psi,\Phi_A)\,.
\end{align}
Splitting the null derivatives as combinations of coordinate basis vectors we can write our system as follows,
\begin{align}\label{hyperbolicityUppercase-Est}
	\p_{T} \mathbf{v}= \mathbf{M}^{p}\p_{p}\mathbf{v}+\mathbf{S}\,,
\end{align}
where $\mathbf{S}$ denotes all the non-principal terms and $\mathbf{M}^{p}$ can be written as,
\begin{align}
	\mathbf{M}^{p} = \begin{bmatrix}
		- s^{\mathring{p}} & 0 & \sg^{\mathring{p}A} \\
		0 & s^{\mathring{p}} & R\sg^{pA} \\
		\frac{1}{2}\sg_A{}^{p} & \frac{1}{2R}\sg_A{}^{p} & 0 
	\end{bmatrix}\,.
\end{align}
Compactifying the coordinates of~\eqref{hyperbolicityUppercase-Est} we find,
\begin{align}
	&J_{T}{}^{\bar{\alpha}}\p_\alpha \mathbf{v} = \mathbf{M}^{p}J_{p}{}^{\bar{\alpha}}\p_{\bar{\alpha}} \mathbf{v} + \mathbf{S}\\
	\Rightarrow&\; \mathbf{X}\p_t \mathbf{v} =  (\mathbf{M}^{p}J_{p}{}^{\bar{p}}-J_{T}{}^{\bar{p}})\p_{\bar{p}} \mathbf{v} + \mathbf{S}\non\,,
\end{align}
where $\mathbf{X}:=J_{T}{}^t-\mathbf{M}^{p}J_{p}{}^t$ and $J_{\alpha}{}^{\bar{\alpha}}$ are the components of the Jacobian associated to the change $(T,R,\theta^A)\rightarrow (t,r,\bar{\theta}^A)$. We can then write that,
\begin{align}\label{hyperbolicityLowercase-Est}
	\p_t \mathbf{v}= \bar{\mathbf{M}}^{\bar{p}}\p_{\bar{p}} \mathbf{v}+\bar{\mathbf{S}}\,,
\end{align}
where $\bar{\mathbf{M}}^{\bar{p}}$ is given by,
\begin{align}
	\bar{\mathbf{M}}^{\bar{p}} = \mathbf{X}^{-1}\mathbf{M}^{p}J_{p}{}^{\bar{p}}\,,
\end{align}
since the Jacobian of the change of coordinates and its inverse are,
\begin{align}
	J^{-1}=\begin{bmatrix}
	1&0& 0 \\
	H'R'&R'&0 \\
	0&0 & 1
	\end{bmatrix},\,J=\begin{bmatrix}
	1&0 & 0 \\
	-H'&\frac{1}{R'}&0 \\
	0&0 & 1
	\end{bmatrix}.
\end{align}
\subsection{Symmetrizer}
At this point we need to find a positive-definite matrix $\mathbf{D}$ such that $\mathbf{D}\mathbf{M}^p$ is symmetric. We know from~\cite{DuaFenGasHil22a} that a choice like,
\begin{align}
	\begin{bmatrix}
		\frac{1}{2}& &  \\
	&\frac{1}{2R^2}& \\
	& & \sg^{AB}
	\end{bmatrix}\,,
\end{align}
works, but for our purposes it is convenient to multiply this matrix with $R^{-2}$, which we can easily see works as well,
\begin{align}
	\mathbf{D}=\frac{1}{R^2}\begin{bmatrix}
		\frac{1}{2}& &  \\
	&\frac{1}{2R^2}& \\
	& & \sg^{AB}
	\end{bmatrix}\,.
\end{align}
Now, changing coordinates to the radially compactified set we can find an equivalent to $\mathbf{D}$ from $\mathbf{D}$ itself. Namely,
\begin{align}\label{H(1+X)-Est}
	\mathbf{D}\mathbf{X} = \mathbf{D}\left(1 + \mathbf{M}^{\Rc}H'\right)\,.
\end{align}
This will be an important step towards finding the energy estimates in what follows.

\section{Estimates for uglies on hyperboloids}\label{Estimates-Est}

\begin{figure}[t!]
\centering
\includegraphics[width=0.5\textwidth]{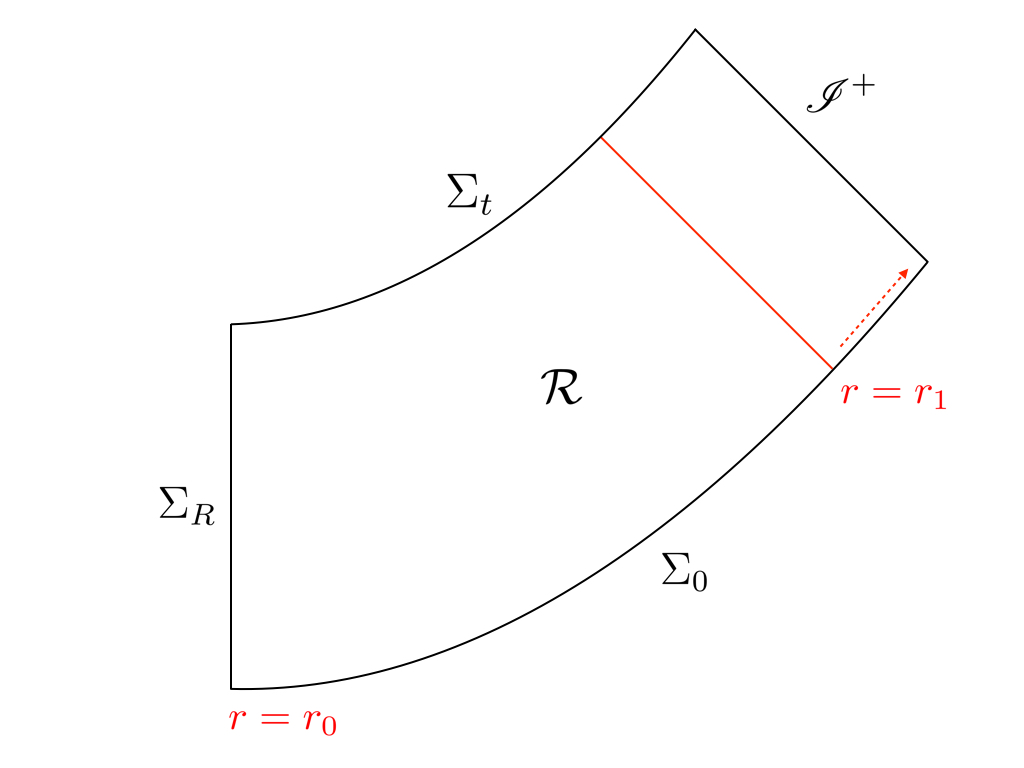}
\caption{A schematic of our geometric setup. We truncate the hyperboloidal slices at $r=r_1$ and then take the limit $r_1\rightarrow r_{\mathscr{I}}$.
  \label{fig:Estimates}}
\end{figure}

Let us consider the spacetime region $\mathcal{R}$ as shown in Figure~\ref{fig:Estimates}, bounded by an initial hyperboloidal slice $\Sigma_0$, a hyperboloidal slice at later time $\Sigma_t$, the boundary of a world-tube $\Sigma_R$ and null infinity $\mathscr{I}^+$.
We define the energy norm on a hyperboloidal slice $\Sigma_t$ as,
\begin{align}\label{EnergyDefinition}
	E^2(t) = \int_{\Sigma_t}\mathbf{v}^T \mathbf{D} \mathbf{v} dV\,,
\end{align}
where $dV$ is written in terms of the spatial coordinates as,
\begin{align}
	dV = R ' R^2drd\Omega= R ' R^2 \sin \bar{\theta} dr d\bar{\theta} d\bar{\phi}\,.
\end{align}
According to~\eqref{EnergyDefinition}, the energy for an ugly equation is then,
\begin{align}\label{Energy}
	E^2(t) = \int_{\Sigma_t}\left( \frac{1}{2R'R^2}\Phi_{\psib}^2+ \frac{2R'-1}{2R'R^4}\Phi_\psi^2\right.\non\\
	\quad\left.+\frac{1}{R^4}\Phi_{\bar{\theta}}+\frac{1}{\sin\bar{\theta}R^4}\Phi_{\bar{\theta}} \right)dV\,.
\end{align}
We temporarily truncate the hyperboloid at $r=r_1$ as shown on Figure~\ref{fig:Estimates} in order to compute the time derivative of the energy and later we take the limit $r_1\rightarrow r_{\mathscr{I}}$, where $r_{\mathscr{I}}$ is the value that $r$ takes as $R\rightarrow \infty$. For now we let this boundary approach the center at the speed of light, meaning that,
\begin{align}\label{ddtr1}
	\frac{dr_1}{dt}=\mathcal{C}_-^r=-\frac{1}{2R'-1}\,.
\end{align}
Naturally, in this setting, the energy varies with $t$ as well as with $r_1$, so we can write that $E(t)=E(t,r_1(t))$. Differentiating $E^2(t,r_1)$ with respect to $t$ we get,
\begin{align}\label{ddtE2}
	\frac{d}{dt}E^2(t,r_1)=\p_t E^2(t,r_1) + \p_{r_1}E^2(t,r_1)\frac{dr_1}{dt}\bigg\vert_{r=r_1}\,.
\end{align}
The first term on the right-hand side can be computed by taking the $t$ partial derivative of~\eqref{Energy} and replacing $\p_t \Phi_{\psib}$, $\p_t \Phi_{\psi}$ and $\p_t \Phi_A$ with the right-hand sides of the field equations. Integrating by parts gives,
\begin{align}
	\p_t &E^2(t,r_1) = \int \frac{1}{2}\left(\frac{\Phi_\psi^2}{R^2}-\Phi_{\psib}^2\right)\bigg\vert_{r=r_0}^{r=r_1}d\Omega \\
	&+ \int_{\mathcal{R}}\frac{1}{R^4\sin\bar{\theta}}\p_A\left[\sin\bar{\theta}\sg^{AB}\Phi_B\left(\Phi_{\psib}+\frac{\Phi_\psi}{R}\right)\right]dV\non\\
	&-\int_{\mathcal{R}}\frac{p}{R^3}\left(\Phi_{\psib}+\frac{\Phi_\psi}{R}\right)^2dV - \int_{\mathcal{R}}\frac{F}{R}\left(\Phi_{\psib}+\frac{\Phi_\psi}{R}\right)dV\,.\non
\end{align}
The second term on the right-hand side is the integral of a total divergence on a 2-sphere and, because there is no boundary, it vanishes entirely. We are then left with,
\begin{align}\label{ptE2}
	\p_t & E^2(t,r_1) = \int \frac{1}{2}\left(\frac{\Phi_\psi^2}{R^2}-\Phi_{\psib}^2\right)\bigg\vert_{r=r_0}^{r=r_1}d\Omega\\
	&-\int_{\mathcal{R}}\frac{p}{R^3}\left(\Phi_{\psib}+\frac{\Phi_\psi}{R}\right)^2dV - \int_{\mathcal{R}}\frac{F}{R}\left(\Phi_{\psib}+\frac{\Phi_\psi}{R}\right)dV\,.\non
\end{align}
We now turn our attention to the second term on the right-hand side of~\eqref{ddtE2},
\begin{align}
	\p_{r_1}E^2&(t,r_1) = \\
	&\lim_{\delta r_1\rightarrow 0}\frac{1}{\delta r_1}\left(\int_{r_0}^{r_1+\delta r_1}-\int_{r_0}^{r_1}\right)\int \mathbf{v}^T \mathbf{D}\mathbf{v} R'R^2d\Omega\,.\non
\end{align}
By the Leibniz's integral rule we get that,
\begin{align}
	&\p_{r_1}E^2(t,r_1) = \int \mathbf{v}^T \mathbf{D}\mathbf{v} R'R^2\big\vert_{r=r_1}d\Omega\\
	&= \int\left(\frac{\Phi_{\psib}^2}{2}+\frac{2R'-1}{2R^2}\Phi_\psi^2+\frac{\Phi_\theta^2}{R^{2-n}}+\frac{\Phi_\phi^2}{R^{2-n}\sin\bar{\theta}^2}\right)\bigg\vert_{r=r_1}d\Omega\,.\non
\end{align}
Putting this together with~\eqref{ddtr1} and~\eqref{ptE2} in~\eqref{ddtE2} and taking the limit $r_1\rightarrow r_{\mathscr{I}}$,
\begin{align}\label{FinalddtE2}
	&\frac{d}{dt}E^2(t)= -\int_{\Sigma_t}\frac{p}{R^3}\left(\Phi_{\psib}+\frac{\Phi_\psi}{R}\right)^2dV\non \\
	&- \int_{\Sigma_t}\frac{F}{R}\left(\Phi_{\psib}+\frac{\Phi_\psi}{R}\right)dV+\int\frac{1}{2}\left(\Phi_{\psib}^2-\frac{\Phi_\psi^2}{R^2}\right)\bigg\vert_{r=r_0}d\Omega\non\\
	&-\int\frac{1}{2}\left(\Phi_{\psib}^2+\frac{\Phi_\theta^2}{R^2}+\frac{\Phi_\phi^2}{R^2\sin\bar{\theta}^2}\right)\bigg\vert_{r=r_{\mathscr{I}}}d\Omega\,.
\end{align}
Note that the fourth integral on the right-hand side contains terms that are proportional to $R^{-2}$ as $R$ goes to infinity. Based on the knowledge about the decay near null infinity of the solutions we found in previous chapters, we would set these terms to zero immediately. However, the point of this chapter is exactly to retrieve that behaviour from energy estimates rather than from heuristic arguments. So we keep these terms for now. Integrating~\eqref{FinalddtE2} with respect to $t$ we find that,
\begin{align}\label{EnergyIntegrated}
	&E^2(t)-E^2(0) =  -\int_{\mathcal{R}}\frac{p}{R^3}\left(\Phi_{\psib}+\frac{\Phi_\psi}{R}\right)^2dVdt \non\\
	&- \int_{\mathcal{R}}\frac{F}{R}\left(\Phi_{\psib}+\frac{\Phi_\psi}{R}\right)dVdt+\int_{\Sigma_R}\frac{1}{2}\left(\Phi_{\psib}^2-\frac{\Phi_\psi^2}{R^2}\right)d\Omega dt\non\\
	&-\int_{\mathscr{I}+}\frac{1}{2}\left(\Phi_{\psib}^2+\frac{\Phi_\theta^2}{R^2}+\frac{\Phi_\phi^2}{R^2\sin\bar{\theta}^2}\right)d\Omega dt\,.
\end{align}
Our aim is to establish upper bounds for the energy $E^2(t)$, so we analyse each of the terms on the right-hand side. The first and fourth integrals are strictly negative, and so can be pulled to the left-hand side or even dropped. In fact we shall keep all those terms except those proportional to $R^{-2}$. The last term of the third integral is also strictly negative and so can be pulled to the left-hand side. We are left with the second integral. We use the Cauchy-Schwarz inequality to find,
\begin{align}\label{EstimateF}
	-\int_{\mathcal{R}}\frac{F}{R}&\left(\Phi_{\psib}+\frac{\Phi_\psi}{R}\right)dVdt \\
	&\leqslant \int_{\mathcal{R}}\left(R^2F^2 + \frac{\Phi_{\psib}}{2R^4}+\frac{\Phi_\psi^2}{2R^6}\right)dVdt\non\\
	&\leqslant \int_{\mathcal{R}}R^2F^2dVdt + C \int E^2(t)dt\,,\non
\end{align}
where $C$ is a positive constant. Plugging~\eqref{EstimateF} into~\eqref{EnergyIntegrated} and using Gronwall's inequality we finally get,
\begin{align}\label{FinalEstimate}
	E^2(t) &+\int_{\mathcal{R}}\frac{p}{R^3}\left(\Phi_{\psib}+\frac{\Phi_\psi}{R}\right)^2dVdt\\
	&+\frac{1}{2}\int_{\mathscr{I}+}\Phi_{\psib}^2d\Omega dt+\frac{1}{2}\int_{\Sigma_R}\frac{\Phi_\psi^2}{R^2}d\Omega dt\non\\
	& \leqslant \left(E^2(0) + \frac{1}{2}\int_{\Sigma_R}\Phi_{\psib}^2d\Omega dt + \int_{\mathcal{R}}R^2F^2dVdt\right)e^{Ct}\,.\non
\end{align}
On the left-hand side we have only the energy squared and strictly positive terms, whereas on the right-hand side we have $E^2(0)$, which is solely determined by initial data on $\Sigma_0$, an integral of $\Phi_{\psib}^2$ on $\Sigma_R$, which must come from assigning boundary data, and an integral of a given function $R^2F^2$ across the entire region $\mathcal{R}$. Note that this estimate reduces to one for the standard wave equation only by setting $p=F=0$. Equation~\eqref{FinalEstimate} is then an upper bound on the energy of solutions to the radially compactified first order form of the ugly equation in flat space.
\section{Conclusions}
\label{section:conclusions}
In this work we showed a deep connection between uglies, bads and goods. Namely, if a given field satisfies an ugly equation with $p=1$, then its rescaled $\psib^a$ derivative must satisfy a good equation in the absence of source terms and a bad equation in their presence. This property of ugly fields was shown for flat space but some generalization of it is expected to carry over to wave operators built from general asymptotically flat metrics. Earlier work suggests that asymptotic solutions to both good and bad fields close to null infinity have logarithms for a large class of initial data, with the difference that those logarithms are linear at every order in the good equation and may appear with higher powers in orders beyond the first in the bad one.  The ugly equation, on the other hand, has been shown to contain no logs at leading order and logs of higher powers at subleading orders. For the case $p=1$ we have shown that the relationship found between these equations is consistent with the asymptotic solutions proven by induction in~\cite{DuaFenGasHil25}. One important consequence of these results is that they allow us to use already existing energy methods for the standard wave equation on Cauchy slices to retrieve information that is valid for the ugly ones, effectively bypassing the process of proving energy estimates for this specific type of equation. Namely we can use well-known recipes to find $L^2$ estimates for the good field and then the Klainerman-Sobolev inequality to show that the rescaled $\psib^a$ derivative of the ugly field has extra decay with respect to the good field. These results are shown specifically for the case $p=1$, which seems to have a more clear relationship to the good equation that the more general $p\neq 1$. \par 
The second part of this work was dedicated to estimating the energy of an ugly equation with any natural $p$ and any given source terms $F$, in radially compactified coordinates on hyperboloidal slices. This proof was made significantly shorter by making use of the tools laid out in~\cite{DuaFenGasHil22a}, where we wrote a first order reduction of an ugly equation, compactified the radial coordinate and showed symmetric hyperbolicity by resorting to a symmetrizer. Here we sketched out this procedure for the flat metric case with stratified null forms as given functions. We computed its total time derivative and integrated in order to find integrals that we could easily estimate through Gronwall's inequality. This allowed us to find an upper bound for the energy of an ugly.\par
The aim of this second part is to be a first step towards establishing energy estimates for the hyperboloidal initial value problem of the regularized compactified first order form of the EFE in GHG as they were presented in~\cite{DuaFenGasHil22a}. Future work along these lines should attempt to generalize these results to include general asymptotically flat metrics, several good and ugly equations and source terms which are allowed to be functions of the evolved fields themselves and their derivatives, as we know is the case with General Relativity.

\acknowledgments

I am grateful to David Hilditch and Edgar Gasper\'in for all the helpful discussions. This work was supported by FCT (Portugal) program PD/BD/135511/2018 and PeX-FCT (Portugal) program 2022.01390.PTDC.

\normalem

\bibliography{Energy_estimates_for_the_good-bad-ugly_model}

\end{document}